\documentclass[preprint,prd,nofootinbib,tightenlines,amsmath]{revtex4}
\usepackage[latin1]{inputenc}
\usepackage{bm}
\usepackage{graphicx}
\usepackage{epsfig}
\usepackage{latexsym}
\usepackage{amsmath}
\usepackage{amsfonts}
\usepackage{amssymb}
\usepackage{color}
\usepackage{hyperref}

\def\de{\partial}
\def\bea{\begin{eqnarray}}
\def\eea{\end{eqnarray}}

\newcommand{\be}{\begin{eqnarray}} % only untightened
\newcommand{\ee}{\end{eqnarray}}

\newcommand{\bmp}{\noindent\begin{minipage}{16cm}}
\newcommand{\emp}{\end{minipage}\vskip 7mm} % 7mm untightened
\def\lsim{\mathrel{\raise.3ex\hbox{$<$\kern-.75em\lower1ex\hbox{$\sim$}}}}
\def\gsim{\mathrel{\raise.3ex\hbox{$>$\kern-.75em\lower1ex\hbox{$\sim$}}}}

\begin{document}

\baselineskip=15pt

\hspace*{\fill} $\hphantom{-}$

\title{Constraints on Conformal Windows from Holographic Duals}
\author{O.~Antipin}
\email{oleg.a.antipin@jyu.fi}
\affiliation{Department of Physics, University of Jyv\"askyl\"a, P.O.Box 35, FIN-40014 Jyv\"askyl\"a, Finland 
\\
and 
Helsinki Institute of Physics, P.O.Box 64, FIN-00014 University of Helsinki, Finland}
\author{K.~Tuominen\footnote{On leave of absence from Department of Physics, University of Jyv\"askyl\"a}}
\email{kimmo.tuominen@jyu.fi}
\affiliation{
CP$^3$-Origins, Campusvej 55, 5230 Odense, Denmark\\
and Helsinki Institute of Physics, P.O.Box 64, FIN-00014 University of Helsinki, Finland\\}

\date{\today}

\begin{abstract} 
\vspace{2mm}
We analyze a beta function with the analytic form of Novikov-Shifman-Vainshtein-Zakharov result in the five dimensional gravity-dilaton environment. We show how dilaton inherits poles and fixed points of such beta function through the zeros and points of extremum in its potential. Super Yang-Mills and supersymmetric QCD are studied in detail and Seiberg's electric-magnetic duality in the dilaton potential is explicitly demonstrated. Non-supersymmetric proposals of similar functional form are tested and new insights into the conformal window as well as determinations of scheme-independent value of the anomalous dimension at the fixed point are presented. 

\end{abstract}

\pacs{PACS numbers: }
\preprint{CP3-Origins-2009-25}
\maketitle

\section {Introduction}

Within the language of quantum field theory some of the most profound insights to the physical realm are encoded into the dependence of the theory on the energy scale. Two basic quantum field-theoretic tools addressing these issues are effective field theories and renormalization which emerged during the theoretical developments in 1970s. Arguably, one of the most basic examples of these phenomena is the evolution of the gauge coupling, given by the beta function. As a well known example of these behaviors, one may consider asymptotically free Yang--Mills (YM) theory where, as the energy scale is decreased from the ultraviolet end towards the infrared, the coupling runs; the renormalization group evolution results in logarithmic increase from near zero values to infinitely strong coupling as the scale of confinement and ultimately non-perturbative physics is reached.

In this work we concentrate on two particular features related to the dependence of a theory from the energy scale which are identified on the level of a generic beta function. One of these is the existence of the fixed point (FP) i.e. a non-trivial zero of the beta function at some finite value of the gauge coupling. Another one is the divergence of the beta function at some finite value of the gauge coupling i.e. existence of a pole in the beta function.

Existence of an infrared FP implies that the theory becomes conformal as we go to the large distances and such theories have received a lot of attention recently in high energy physics due to the progress made in understanding the phase diagram of strongly coupled gauge theory as a function of number of colors and flavors as well as matter representations \cite{Sannino:2004qp,Dietrich:2006cm}. Phenomenologically these theories are important for beyond Standard Model (SM) model-building within the frameworks of walking technicolor \cite{Dietrich:2005jn} and unparticles \cite{Sannino:2008nv}.

On the other hand, if beta function has a pole, the evolution of the gauge coupling will cease at a finite value of the coupling as we approach the pole scaling from the ultraviolet by integrating out high energy degrees of freedom. 

Since both the presence of the pole as well as a fixed point in the beta function may be scheme dependent properties, one may ask how much physical significance can be given for, say, a pole which can be present in one scheme but absent in another. This question is a basic motivation for the study on which we now embark. 

For the four dimensional quantum field theory we will at first concentrate on the known exact results for the beta function in certain quantum field theories, where we may see both of the above mentioned features. Concretely, we consider the exact $\beta$-function of supersymmetric QCD (SQCD), first derived by Novikov-Shifman-Vainstein-Zakharov (NSVZ) \cite{Novikov:1983uc},
\be \label{NSVZ}
	\beta (\lambda) &=& - 2\lambda^2
	\frac{\beta_0(Y)+r\gamma(\lambda)}{1-j\lambda} \ , \\
	\gamma(\lambda) &=& - 4\lambda\frac{C_2(R)}{N_c} + {\mathcal{O}}(\lambda^2) \ ,
\ee
where $r= Y \equiv N_f/N_c$ and $N_{f}$ $(N_c)$ is a number of flavors (colors), $j=2$, $\beta_0= 3-Y$ is the first beta function coefficient and $C_2(R)$ is the quadratic Casimir of the gauge group. Furthermore, $\gamma(\lambda)$ is the anomalous dimension of the matter superfield with the leading perturbative term shown in the above equation and $\lambda=\frac{g_{YM}^2 N_c}{(4\pi)^2}$ is the running 't Hooft coupling.

Unfortunately, corresponding results for non-supersymmetric theories are not known and progress in this direction is typically of conjectural nature only. In this regard, proposal for the beta function of non-supersymmetric theories with arbitrary fermionic matter content inspired by NSVZ form appeared in \cite{Ryttov:2007cx}. Replacing anomalous dimension with its leading perturbative expression we arrive at the following symbolic form of the beta function parametrizing both supersymmetric and non-supersymmetric cases:  
\be
\label{betaFP}
	\beta(\lambda) &=&- 2 \beta_0 \lambda^2 \frac{(1-v \lambda)}{1- j \lambda } \ ,
\ee
where in SQCD case coefficient $v$ can be read off from Eq. (\ref{NSVZ}) while corresponding coefficients $j$, $\beta_0$ and $v$ for the non-supersymmetric conjecture will be specified later as we discuss concrete examples.

Another approach towards understanding of strongly coupled gauge theories is based on the celebrated AdS/CFT correspondence \cite{Maldacena:1997re,Witten:1998qj} allowing to mathematically relate certain large $N_c$ gauge theories to string theory at certain limit. Even though the most presice version of this holographic correspondence exists for the four-dimensional (4D) $\mathcal{N}=4$ Super YM (SYM) theory, which is quite distant from the real world, extended holography was tested in many environments giving encouraging results \cite{Polchinski:2001tt,Erlich:2005qh}. In particular, an interesting approach based on the coupled five dimensional (5D) dilaton-gravity system was developed recently in \cite{Gursoy:2007cb,Gursoy:2007er} allowing to encode exact beta function of QCD-like theories directly into the structure of the potential for the dilaton field. For applications of this method most closely related to our work see \cite{Jarvinen:2009fe,Alanen:2009xs,Alanen:2009ej,Pirner:2009gr,Alvares:2009hv}.  

In this paper we will show how the vacuum solution to the coupled 5D dilaton-gravity system, completely specified by the dilaton potential, inherits poles and fixed points of the beta function of a 4D boundary theory given by Eq. (\ref{betaFP}). Our exemplary core case study will be given by SYM-SQCD theories taking advantage of the complete non-perturbative knowledge of the supersymmetric gauge theory on the boundary. Sensitivity of the dilaton potential to the particular renormalization scheme used to arrive at the complete beta function will be tested via rewriting Eq. (\ref{betaFP}) in a variety of different schemes. In the SQCD case, we will see how the physical picture of the electric-magnetic duality encoded into the structure of the dilaton potential will emerge. Of course it is clear that for these cases there is nothing further to learn about the field theory from the gravity solution; rather we see these examples as consistency checks on the properties of the gravity solution. These insights will turn out useful, since ultimately we will discuss similar physical results in a non-supersymmetrical proposals with the structure of Eq. (\ref{betaFP}). We will show that here the gravity duals may give new insights to these proposals. As particular applications we consider predictions for the lower end of the conformal window (CW) \cite{Poppitz:2009uq,Poppitz:2009tw,Armoni:2009jn} as well as scheme-independent values of the anomalous dimension at the FP.

In the next section we briefly review the holographic framework we are going to use. We then test this method with a detailed examples of SYM-SQCD in Sec.\ref{SQCD} and further apply it to the non-supersymmetric beta functions proposals in Sec.\ref{RS}. Our conclusions and outlook are given in Sec.\ref{outlook}.

\section{The Holographic Method}\label{setup}

As is well known, in the low energy limit string theory can be approximated by pure supergravity, which, in turn, is expected to give an adequate qualitative holographic description of a pure YM in a large $N_c$ limit. Briefly, the steps in a top-down approach to the dual description of the pure YM can be summarized as follows \cite{Gursoy:2007cb,Gursoy:2007er}: In a first step an effective action for the long range fields of the $D_3$ branes in a IIA/IIB string theory is obtained. The minimal degrees of freedom at this step include the gravitational fields (dilaton and the axion), as well as the five-form flux controlling the number of $D_3$ branes. To simplify the theory further, axion field can be neglected in the leading order and the string frame action is transformed to the Einstein frame. From this Einstein frame action, the equations of motion for the five-form are solved and inserted back to the action. Finally, integrating over the $S_5$-sphere, the action for the coupled 5D dilaton-gravity theory is
\be\label{action-einst}
S = {M^3 N_c^2}\int d^5 x \sqrt{-g}\left[ R - {4\over 3}
 g^{\mu\nu}\de_\mu \phi \de_\nu\phi + V(\phi)\right]\,.
\ee 
where $M$ is the five-dimensional Planck scale and $N_c$ is the number of colors.

Working in the so called domain-wall coordinates, we assume that the above action is minimized by a background of the form
\be
\label{sol} 
g_{\mu\nu} = du^2 + e^{2A(u)}\eta_{ij} dx^i dx^j, \quad \phi = \phi(u),
\ee
where $x^i$ are the usual 4D spacetime coordinates of the Minkowski space with metric $\eta_{ij}={\rm{diag}}(-,+,+,+)$.

The dilaton potential $V(\phi)$ is expected to be a non-trivial function which is in a one-to-one correspondence with the $\beta$-function of the boundary gauge theory. The independent Einstein's equations take the following form:
\be\label{einsteq}
(\phi^\prime(u))^2 = -{9\over 4} A^{\prime\prime}(u), \quad V(\phi) = 3 A^{\prime\prime}(u) + 12 (A^{\prime}(u))^2.
\ee

These equations can  be written in the first-order form in terms of a superpotential $W(\phi)$ as
\bea
&&\phi^\prime(u) = {d W\over d\phi}, \qquad  A^\prime(u) = -{4\over 9} W, \label{einsteinsuper}\\
 && V(\phi) =
-{4\over 3}\left({d W\over d\phi}\right)^2 + {64\over 27} W^2.\label{VtoW}
\eea

The 4D 't Hooft coupling $\lambda=g_{YM}^2 N_c/(4\pi)^2$ is identified with the dilaton field $\lambda=  e^{\phi}$ and familiar holographic interpretation 
of 4D energy scale $E$ with the scale factor $A(u)$, $\log E \leftrightarrow A(u)$ is used. With these identifications, it follows that the $\beta$-function of the 't Hooft coupling is related to 5D fields as:
\be
\label{beta}
\beta(\lambda) \equiv {d \lambda \over d \log E} = \lambda {d \phi \over d A}.
\ee

Generally the relation between $\log E$ and $A$ is some complicated function and the identification $\log E=A$ holds only in the asymptotic weak coupling limit. Also the identification of the field theory coupling $\lambda$ with the bulk dilaton $\phi$ requires the asymptotic behavior $\beta(\lambda)=-b_0\lambda^2-b_1\lambda^3+\dots$. Definition of the field theory coupling $\epsilon(\lambda)$ satisfying $\beta(\epsilon)=-b_0\epsilon^2-b_1\epsilon^3+\dots$ in the weak coupling limit, and a functional relation between $\log E$ and $A$ defines a renormalization scheme for the field theory. In the following section we will introduce several different schemes and study the resulting behaviors in the gravity solution. For all schemes we consider we fix $\log E=A$. 

We will mostly use variable $X$, introduced in \cite{Gursoy:2007cb,Gursoy:2007er}, which is directly related to the logarithmic derivative of $W(\phi)$: 
\be\label{x}
 X = -\frac{3}{4}{d\ln W\over d\phi} = \frac{\beta(\lambda)}{3\lambda}.
\ee

It is a simple exercise to combine Eqs. (\ref{VtoW}) and (\ref{x}) to write the dilaton potential as:
\be
\label{potential}
V(\lambda)= \frac{64}{27} W^2(\lambda) \left[1-X^2(\lambda)\right].
\ee

Clearly, zeros of the dilaton potential, if they exist, will be encoded into the function $1-X^2$ provided that the superpotential $W$ has no zeros. This will always be the case in this paper as can be seen explicitly for each concrete example considered in what follows. Thus, if $|\beta(\lambda)|>3\lambda$ for some physical $\lambda$, dilaton potential will change sign.

At the Gaussian FP, $|\beta(\lambda)|/(3\lambda)\sim \lambda \to 0$ and, thus, as $1-X^2 \to 1$, dilaton potential $V \to 64 W^2(0)/27 >0$.
Also, generally, if $\beta$-function has a pole and does not have non-trivial FP's, $|\beta(\lambda)|>3\lambda$ will hold at some point and we have a sign change in the dilaton potential before we approach the pole. Hence, generally, if there is a pole in the beta function we expect to find a zero in the dilaton potential.

On the other hand, the FP in the beta function appears in the dilaton potential as an extremum. From (\ref{x}) and (\ref{potential}) one easily derives that
\be
\frac{dV}{d\lambda}=\frac{128}{27}W^2(\lambda)\left(-\frac{4}{3\lambda}(1-X^2(\lambda))-\frac{dX(\lambda)}{d\lambda}\right)X(\lambda).
\ee
Therefore, if $\beta(\lambda^\ast)=0$, Eq. (\ref{x}) implies that $X=0$ which further implies an extremum of $V(\lambda)$. Note however, that the extremum of the dilaton potential is only a necessary but not sufficient condition for the FP.

In relation to these considerations we also note that one important class of asymptotes discussed in \cite{Gursoy:2007cb,Gursoy:2007er} has $X(\lambda) \to -1/2$ as $\lambda\to +\infty$ (working in the variables where this limit exists) and thus dilaton potential might never change a sign. For this to happen, additional change in the curvature of the beta function is required because beta function always needs to stay above the $\beta(\lambda) = - 3\lambda$ line while unambiguous UV limit for asymptotically free gauge theory in any coupling scheme requires it to asymptote to $\beta(\lambda)\approx -\beta_0\lambda^2-\beta_1\lambda^3$. 

As a final remark on the method we note that although we will consider theories with matter fields, we do not embark on the detailed study of adding flavors using probe branes in the bulk. Rather, we consider including the anomalous dimension as a function of the coupling constant into the beta function and assume this to give adequate effective description of fermions at least in the domain where the coupling is small. In order to have some control over this assumption we will begin with the example of SYM and SQCD theories where we can verify the implications of the gravity solution against the known results for the corresponding field theory.

\section{Poles and fixed points: SYM-SQCD case study}\label{SQCD}

In this section we will see how the dilaton potential inherits the physics of the beta functions parametrized as in Eq. (\ref{betaFP}). First, we deal with beta function without FP (only pole) which amounts to $v=0$ in Eq. (\ref{betaFP}). This case will be illustrated with a pure SYM example. Then, we will allow for the FP using the leading perturbative term of the anomalous dimension. This case will be illustrated by SQCD example with a detailed analysis of the conformal window and Seiberg duality \cite{Seiberg:1994pq,Intriligator:1995au}. 

We start with Eq. (\ref{betaFP}) where the parameters are $\beta_0 = 3-Y$, $j=2$ and $v=2Y/\beta_0$ with $Y\equiv N_f/N_c$. Also, for any number of flavors the coefficient $j$ is fixed to the pure SYM value $j=2$. 

\subsection{No FP ($Y=v=0$): SYM case study \label{Poles1}}
 
Let us start by considering the coupling constant $\lambda$ and the beta-function given by (\ref{NSVZ}). Direct use of Eq. (\ref{betaFP}) gives
\be\label{direct}
1-X^2 = 0 = 1- \left(\frac{2\beta_0\lambda}{3(1-j\lambda)}\right)^2 \Longrightarrow \lambda=\frac{1}{j\pm \frac{2}{3}\beta_0} = \frac{1}{4},\infty,
\ee
while from Eq. (\ref{x}) the superpotential reads
\be\label{W3}
W=W_0(1-j\lambda)^{-\frac{8\beta_0}{9j}}.
\ee
Here integration constant $W_0=W(\lambda=0)=9/(4\ell)$ is related to the asymptotic AdS length $\ell$. The integration constant $W_0$ will only serve to normalize the results; in our numerical examples we will fix $W_0=1$. We observe that in terms of variable $\lambda$ the spacetime ends at $\lambda=1/j=1/2$ where the superpotential becomes complex and the zero of the dilaton potential occurs at $\lambda=1/4$, i.e. ``half-way'' towards the pole.

Next, consider a scheme change which brings the beta function of Eq. (\ref{betaFP}) to the one-loop exact form $\beta(\lambda_h)=-2\beta_0 \lambda_h^2$ through the definition of  the new coupling $\lambda_h$ via $\lambda_h^{-1}= \lambda^{-1} + j \ln(\lambda)$. This is the famous Shifman-Vainstein holomorphic coupling \cite{Shifman:1986zi} (see also \cite{ArkaniHamed:1997mj}). Scheme change is one-to-one for the values of $\lambda$ from zero to the pole value 1/2. We further identify the dilaton with the holomorphic coupling $\lambda_h=e^{\phi}$ as otherwise we need $\lambda(\lambda_h)$ inverse transform which cannot be expressed in terms of elementary functions and therefore is not very illuminating. We have again a zero in the dilaton potential,
\be
1-X^2 = 0 = 1- \left(\frac{2\beta_0}{3}\lambda_h\right)^2 \Longrightarrow \lambda_h=\pm \frac{3}{2\beta_0}= \pm\frac{1}{2},
\ee
 and the superpotential in this case reads
\be\label{W1}
W=W_0 \text{exp} \left[\frac{8\beta_0\lambda_h}{9}\right].
\ee

Note also that the relation between the coupling and the scale factor $A(\phi)$ is very transparent in these variables: Starting from the infinitesimal form of Eq. (\ref{beta}), we first obtain
\be
d\left(\frac{1}{\lambda_h}\right)=2\beta_0 dA,
\ee
which implies that 
\be\label{A}
A= \frac{1}{2\beta_0\lambda_h} + C.
\ee
Now, holomorphic coupling reaches its maximum value $\lambda_h^{max}=1/(2 - 2 \log (2))$ for $\lambda=1/2$. Fixing $A(\lambda_h^{max})=0$, we interpret the scale factor as diverging in UV where $\lambda_h\to 0$ and shrinking to zero in the infrared, i.e. when approaching the pole in the original $\lambda$-variable.

\begin{figure}[htb]
\centering
\includegraphics[width=3.0in]{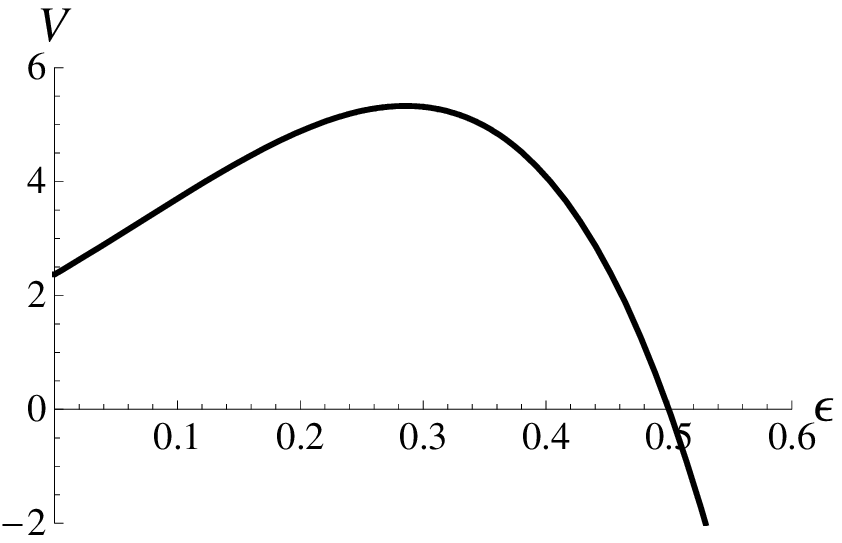} 
\includegraphics[width=3.0in]{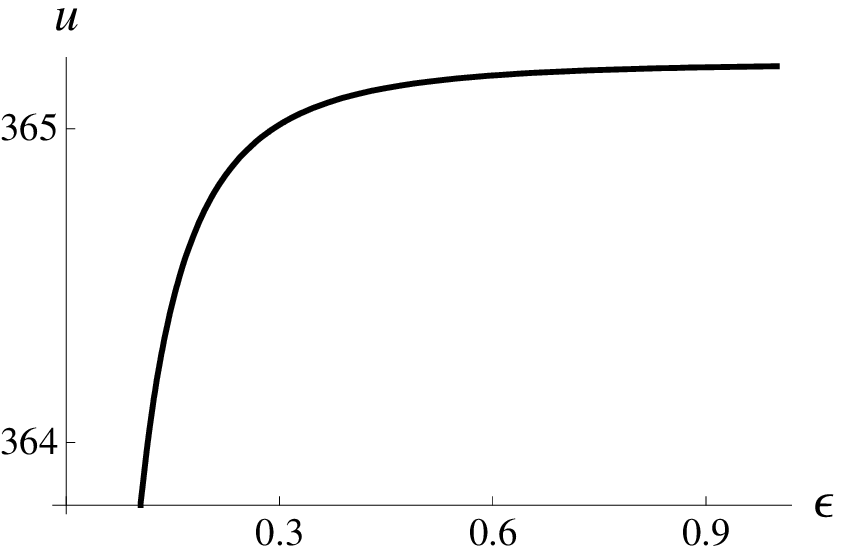}
\caption{(left) Dilaton potential in the two-loop exact scheme when the dilaton is $\lambda=\exp(\phi)$. (right) Illustration of the running of the coupling for the scheme corresponding to the dilaton potential in the left panel. The value $W_0=1$ was used for both graphs.}
\label{pot}
\end{figure}

After these preliminary considerations we introduce a change of variables which will be central to our study. We define the coupling constant 
\be
\epsilon = \lambda/(1-j\lambda),
\ee
and the resulting $\beta$-function is
\be
\beta(\epsilon)=-2\beta_0 \epsilon^2 (1+j\epsilon).
\ee 
In other words, we defined a scheme which is two-loop exact. The above beta function has a FP at the unphysical value of the coupling, $\epsilon^*=-1/j=-1/2$. The scheme change implied by this change of variables is one-to-one everywhere and maps $\lambda\in[0,1/2)$ onto the interval from zero to infinity. Consider first keeping the dilaton unchanged, and define $\lambda=e^{\phi}$. Then we have
\be
1-X^2 = 0 = 1- \left(\frac{2\beta_0}{3}\epsilon\right)^2 \Longrightarrow \epsilon=\pm \frac{3}{2\beta_0}= \pm\frac{1}{2}.
\ee
For completeness we present again also the superpotential,
\be\label{W2}
W=W_0(1+j\epsilon)^{\frac{8\beta_0}{9j}}, 
\ee
and plot the full dilaton potential using Eq. (\ref{potential}) in the left panel of Fig.\ref{pot}. Note that the superpotential (\ref{W2}) is obtained from (\ref{W3}) by changing variables from $\lambda$ to $\epsilon$, which is a consequence of the fact that the dilaton is defined in both cases using the coupling $\lambda$. 

Furthermore, it is trivial to find the transformation to the domain-wall variable $u$ using first equation in Eq. (\ref{einsteinsuper}). We find that:
\be \label{coordtr}
u(\epsilon)-u(s)=\frac{1}{W_0 d}\int_s^{\epsilon} \frac{(1+jt)^{-\frac{d+j}{j}}}{t^2} dt,
\ee
where we introduced shorthand $d\equiv\frac{8}{9}\beta_0$.
The integrand diverges at the lower limit $s \to 0$ corresponding to the UV limit in the domain-wall variable $u\to -\infty$. We plot Eq. (\ref{coordtr}) in the right panel of Fig.\ref{pot} for $s=10^{-3}$. We observe that $\epsilon$ diverges at the finite value of $u$ confirming the absence of any FPs in the boundary gauge theory. 

Of course nothing forbids to take $\epsilon=e^{\phi}$  in the above two-loop exact scheme, and qualitatively nothing changes. The dilaton potential will have a zero given by the solution of
\be \label{4roots}
1-X^2 = 0 = 1- \left(\frac{2\beta_0\epsilon}{3}(1+j\epsilon)\right)^2 \Longrightarrow \epsilon=\frac{-\beta_0\pm\sqrt{\beta_0(\beta_0\pm 6j)}}{2j\beta_0} \text{(four roots)}.
\ee
For SYM ($\beta_0 = 3$, $j=2$), out of the four solutions the only real and positive one corresponds to the choice of both $+$ signs in Eq.\ref{4roots} and is $\epsilon_1^* \approx 0.309$. Finally, the superpotential is given by
\be\label{W4}
W=W_0 \text{exp} \left[\frac{8\beta_0\epsilon}{9}\left(1+\frac{j}{2}\epsilon\right)\right].
\ee

Even though we used different variables and different identifications for the dilaton, the common property is that the dilaton potential has a zero. We further note, that superpotentials in Eqs. (\ref{W2}) and (\ref{W4}) pass the criteria for confinement presented in \cite{Gursoy:2007er}. There it was shown that class of superpotentials growing faster than $($log $\epsilon)^{P/2} \epsilon^{2/3}$ as $\epsilon \to \infty$ (for some $P\ge 0$) correspond to a confining theory. In our case we have $P=0$ and $W\sim \epsilon^{4/3}$ for $\epsilon \to \infty$ in (\ref{W2}) while the superpotential in (\ref{W4}) grows exponentially. Physical results are expected to be scheme independent and hence also (\ref{W3}) and (\ref{W1}) correspond to confining potentials. The two-loop exact scheme we considered is particularly useful due to the fact that $\epsilon$ changes from $0$ to $\infty$ which eases the comparison with \cite{Gursoy:2007er}; all other schemes lead to physically equivalent results as we have seen. 

\subsection{FP and the poles: SQCD case study \label{FPP} }

With the notation and concepts established in previous subsection we now move on to discuss theories with matter. As a concrete example we consider SQCD as a benchmark where FP and Seiberg duality map \cite{Seiberg:1994pq} $Y \to \frac{Y}{Y-1}$ exist for $\frac{3}{2} < Y < 3$. Also, $v(Y=3/2)=2$ and $v(Y=3)=\infty$. The superpotential and the dilaton potential can be easily calculated as direct generalization of the results in the previous subsection. 

In its original form (\ref{NSVZ}) has support only for the values of the coupling $\lambda$ on the interval from zero to $1/j=1/2$; note that this pole value is independent of matter fields. To better compare against literature we again consider redefinitions of the coupling. The holomorphic coupling is not very illuminating in this case as, regardless of the dilaton definition, we inevitably need $\lambda(\lambda_h)$ inverse transform which cannot be expressed in terms of elementary functions. We escaped this problem when considering the holomorphic scheme in Sec.\ref{Poles1} due to the absence of $\sim v\lambda^3$ term in the numerator of the original beta function. 

The two-loop scheme, $\epsilon = \lambda/(1-j\lambda)$ which implies $\beta(\epsilon)=-2\beta_0 \epsilon^2 (1+(j-v)\epsilon)$, turns out to be very convenient for us here. We will also exclusively use the identification $\lambda=\exp(\phi)$ from now on. Observe first that this beta function has FP at $\epsilon^*=-1/(j-v)$ and this FP appears on the positive real axis as lower boundary of the CW at $Y=3/2$ ($v=2$) is reached from below. In the dilaton potential the fixed point $\epsilon^*$ corresponds to the extremum point $V^{\prime}(\epsilon^*)=0$. As we move across the CW from lower to the upper boundary, this FP will move from $\infty$ to the zero value of the coupling. Notice that under Seiberg's duality map, $Y \to \frac{Y}{Y-1}$, we have $\epsilon^* \to \frac{1}{\epsilon^*}$, as expected. 

\begin{figure}[htb]
\centering
\includegraphics[width=3.0in]{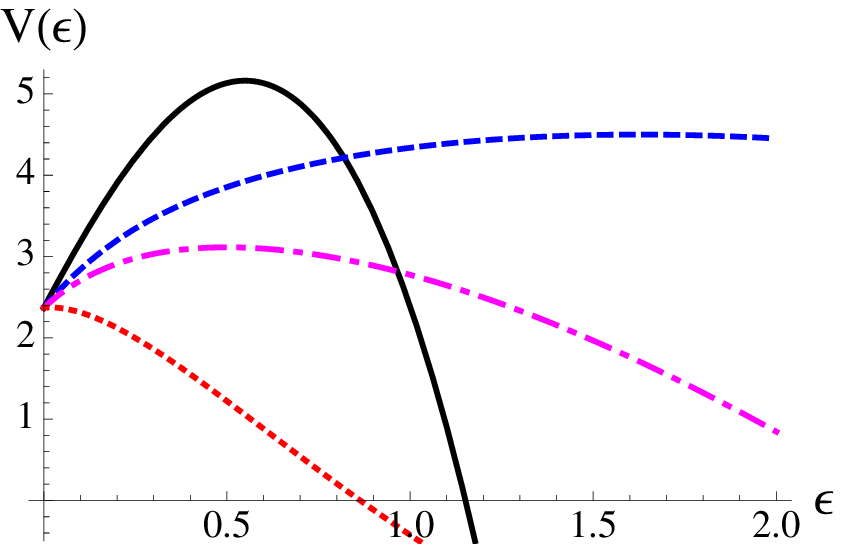}
\includegraphics[width=3.0in]{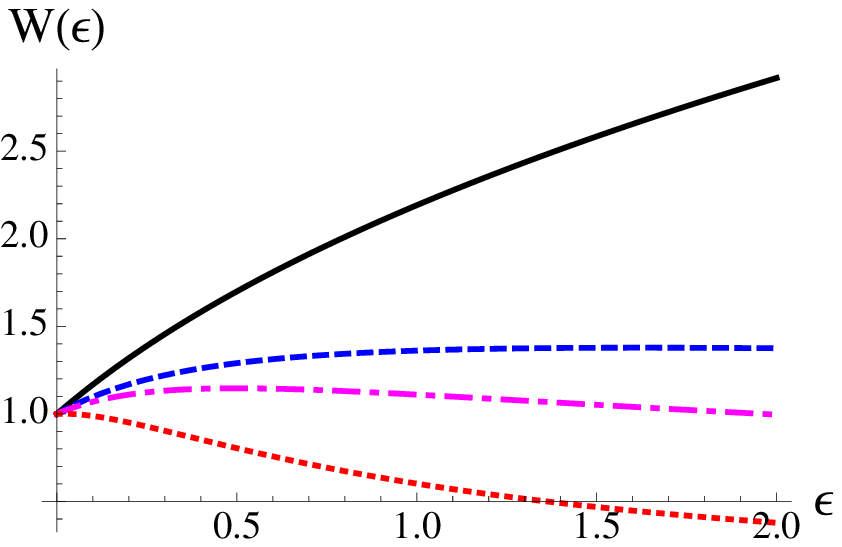} 
\caption{Left panel: The dilaton potential $V(\epsilon)$ for values $Y=1$ (solid), $Y=1.7$ (dashed), $Y=2$ (dash-dotted) and $Y=2.9$ (short dashes). The normalization factor $W_0$ has been set to unity. Right panel: Corresponding superpotentials. The notation is the same as for the figure in the left panel.}
\label{vw_pots}
\end{figure}

With these variables we have 
\be
X=\frac{2\beta_0\epsilon(1+(j-v)\epsilon)}{3(1+j\epsilon)},
\ee
and in Fig. \ref{vw_pots} we illustrate the qualitative features of the dilaton potential and superpotential as the conformal window, starting at $Y=3/2$, is entered from below. Below the conformal window the theory is confining as the rapid growth of the superpotential as a function of the coupling shows, and the coupling diverges in the infrared as can be explicitly verified via (\ref{coordtr}). On the other hand, above the lower boundary the behaviors of both the dilaton potential and the superpotential become markedly different. As we are within the conformal window, confinement is lost as is shown by the superpotential which in this region is a monotonously decreasing at large values of the coupling. The extremum of the dilaton potential corresponds to a physical fixed point moving from large coupling towards zero as $Y$ increases from 3/2 to 3 where asymptotic freedom is lost at the upper boundary of the conformal window. Note, that the dilaton potential within the conformal window has zero determined by
\be
1-X^2 = 0 = 1- \left(\frac{2\beta_0\epsilon_0(1+(j-v)\epsilon_0)}{3(1+j\epsilon_0)}\right)^2.
\label{twoloop1}
\ee
The value of $\epsilon_0$ is always larger than the fixed point value $\epsilon^\ast$ and in particular $\epsilon_0$ will remain nonzero as $\epsilon^\ast\rightarrow 0$ at the upper boundary of the conformal window.

To explicitly verify that within the conformal window the above potentials correspond to nonconfining gauge theory on the boundary we again recall the results of \cite{Gursoy:2007er}. In terms of the variables we are using here, confinement corresponds to superpotentials growing faster than $\epsilon^{2/3}/(1+2\epsilon)^{2/3}$, and we dropped here the possible positive power of $\log\epsilon$ which do not arise in our potentials anyway. At large $\epsilon$ the leading behavior is
\be
W(\epsilon)/[\epsilon^{2/3}/(1+2\epsilon)^{2/3}]\sim \epsilon^{-\frac{8Y}{9}+\frac{4}{3}}+{\mathcal{O}}(\epsilon^{-\frac{8Y}{9}+\frac{1}{3}}),
\ee
which implies that the change from confining to nonconfining theory occurs at $Y=3/2$ consistently with Seiberg's analysis.

\begin{figure}[htb]
\begin{minipage}[c]{0.5\linewidth}
%\centering
\hspace{-1cm}\includegraphics[width=3.2in]{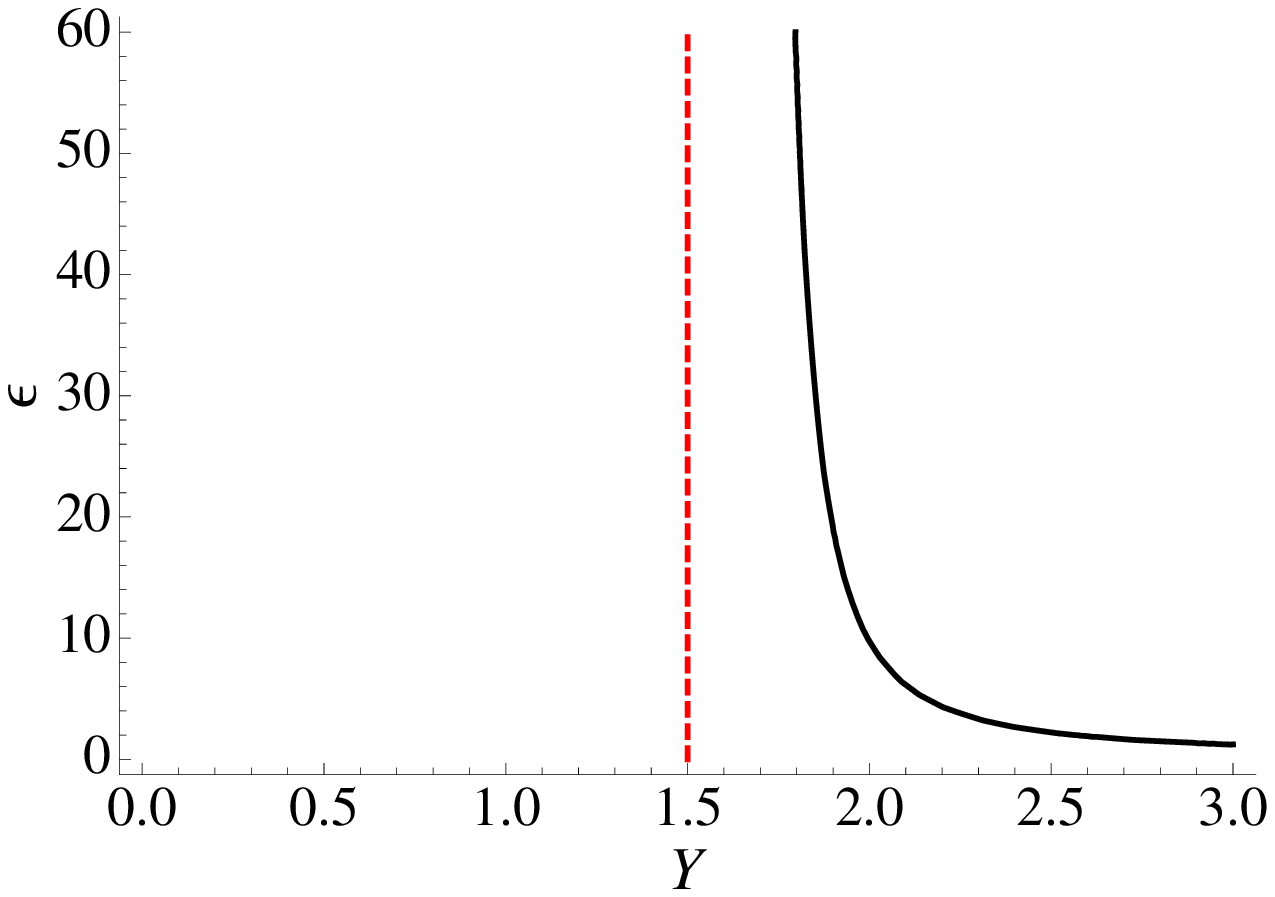} 
\end{minipage}%
\begin{minipage}[c]{0.5\linewidth}
\centering
\includegraphics[width=3.2in]{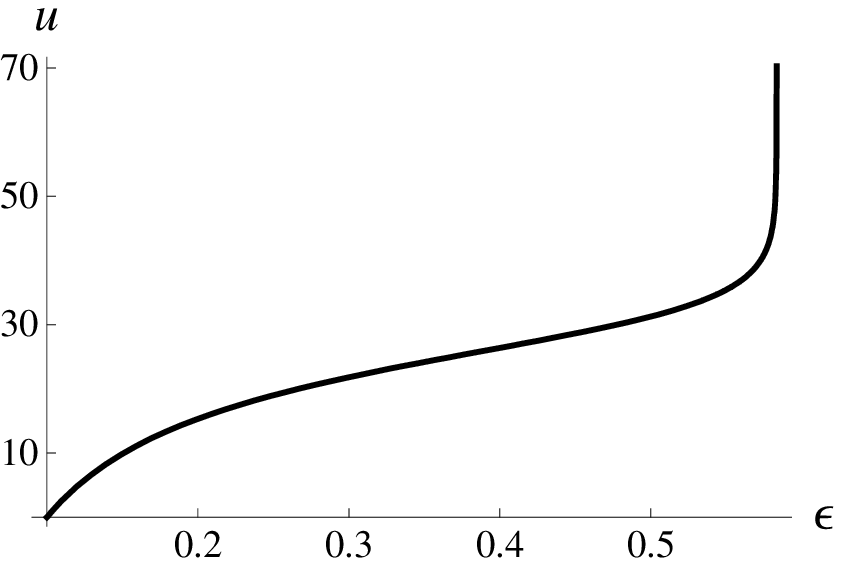} 
\end{minipage}
\caption{Left panel: The solid curve shows value of $\epsilon(Y)$ where the superpotential $W$ crosses below the limit in (\ref{conf_limit}). The dashed vertical line shows the location of the lower boundary of the conformal window. Right panel: Illustration of the FP in the SQCD corresponding to $Y=1.95$ (for which $\epsilon^\ast=0.583$). As for the other figure, also here $W_0=1$. }
\label{FP_pot}
\end{figure}

For illustration, in the left panel of Fig.\ref{FP_pot} we plot the curve $\epsilon(Y)$ corresponding to values of $Y$ for which the inequality
\be
W(\epsilon)\le  \epsilon^{2/3}/(1+2\epsilon)^{2/3}
\label{conf_limit}
\ee
is satisfied for $\epsilon\ge\epsilon(Y)$; the boundary theory becomes nonconfining as the conformal window is entered at $Y=3/2$.

Finally, within the conformal window the existence of the fixed point can be confirmed by looking at the evolution of the coupling in the infrared as illustrated in Fig. \ref{FP_pot} for $Y=1.95$ for which we expect the appearance of the FP at $\epsilon^* \approx 0.583$. We observe that $\epsilon$ indeed flows to this FP at low energies.

We emphasize that in Eq.(\ref{betaFP}) we included anomalous dimension perturbatively and thus, in principle, we expect our approximations to work only in the vicinity of the upper boundary of the CW. However, we have seen that this approximation leads to interesting and intuitive picture encoded into the structure of the dilaton potential across the {\it whole} CW. Moreover, the analysis we discussed above is fully consistent with Seiberg's analysis of the conformal window within the boundary theory.  

To conclude this section, we have seen that the gravity dual obtained using the beta-function of supersymmetric QCD exactly reproduces the known behavior of the field theory. Of course for the supersymmetric case everything is known without the gravity solution. However, this section serves as an important testbench for the holographic description of conformal window as in the next section we aim to apply these results in the nonsupersymmetric case. 

\section{Estimates for the conformal window in non-supersymmetric theories}\label{RS}

Encouraged by the results in the supersymmetric theories, we now deal with the non-supersymmetric SU$(N_c)$ gauge theories with fermionic matter in some representation $R$ of the gauge group. As a concrete case study we consider the Ryttov-Sannino conjecture for the beta function, which was in turn inspired by the exact NSVZ result \cite{Ryttov:2007cx}. Consequently, non-supersymmetric beta function will be given by Eq. (\ref{NSVZ}) where the first beta function coefficient is $\beta_0 =(11-2Z)/3$ with $Z\equiv 2T(R)N_f/N_c$. We generalized slightly our notation from supersymmetric cases of previous section, $Y\rightarrow Z$, to allow for the matter in arbitrary representation of the SU$(N_c)$ gauge group. We are using standard group theory notations with conventions adopted in \cite{Sannino:2008ha}. Trivially, for the fundamental matter $Z=Y$ and we recover the notation used up to this point. The well-known scheme independent second beta function coefficient is 
\be \label{beta1}
\beta_1 =\frac{34-10Z}{3} - \frac{2 C_2(R) Z}{N_c}.
\ee
The higher coefficients are scheme dependent and will not be considered here. For fundamental matter in the large $N_c$ limit we have $\beta_1=\frac{34-13Y}{3}$. 

Unlike for supersymmetry, there is no principle dictating the unique assignment of $j$ and $r$ coefficients in Eq. (\ref{NSVZ}) for non-supersymmetric case. However, there are three immediate constraints on any given proposal: First, we need to reproduce the scheme-independent perturbative two-loop result once expanding Eq. (\ref{NSVZ}) to the ${\mathcal{O}}(\lambda^3)$ order. Second, for pure YM we can compare to high precision data on the running of the coupling constant. Third, if we consider the theory with single Weyl fermion transforming in the adjoint representation of the gauge group the theory (in the large $N_c$ limit) should correspond to SYM for which exact results are known. 

Now, we again take anomalous dimension into account perturbatively
\be \label{gam}
\gamma(\lambda) = 6\frac{C_2(R)}{N_c} \lambda + {\mathcal{O}}(\lambda^2),
\ee
which amounts to 
\be\label{dict}
r=-\frac{\beta_0 N_c}{6C_2(R)} v
\ee
dictionary between Eq. (\ref{NSVZ}) and Eq. (\ref{betaFP}) in this case. Let us now see how the three constraints mentioned above appear. Once expanding Eq. (\ref{betaFP}) to the ${\mathcal{O}}(\lambda^3)$ order we have $\beta(\lambda)=-2\beta_0 \lambda^2 (1+(j-v)\lambda)+ \dots$ . To take into account the first constraint, we note that it is only $j-v$ combination which parametrizes the unique second beta function coefficient, i.e. $\beta_1=\beta_0 (j-v)$. Clearly there are infinite number of choices for $j$ and $v$ which satisfy this equation. Then consider the second constraint: Given the fit performed in \cite{Ryttov:2007cx}, we require that
$j=\beta_1$/$\beta_0=34/11$ in the pure YM limit \cite{jnote}. In other words, the part of $\beta_1$ which is independent of $N_f$ and fermion representations is contained entirely in $j$. Finally, let us consider the third constraint. This applies only to a single adjoint Weyl fermion for which $\beta_0=3$ and $\beta_1=6$, so that from the two-loop matching we have $j-v=2$. Now we consider the ansatz (\ref{NSVZ}) for single adjoint Weyl fermion,
\be
\beta(\lambda)=-2\lambda^2\frac{3-\frac{v}{2}\gamma_{\rm{Adj}}}{1-j\lambda},
\ee
and, assuming that the schemes coincide, equate this with the corresponding NSVZ result for SYM which is
\be
\beta_{SYM} (\lambda) = - 2\lambda^2 \frac{\beta_0}{1-2\lambda}=-\frac{6\lambda^2}{1-2\lambda}.
\ee
The resulting equation is used to solve for $\gamma_{\rm{Adj}}$. Demanding that this reproduces the known result,
\be
\gamma= \frac{6\lambda}{1-2\lambda},
\ee
leads to the relation $j-v=2$ which coincides with the relation obtained from two-loop matching.
Note that this matching procedure implies that $\lambda(\mu)\langle\bar{\psi}\psi\rangle$ is renormalization group invariant as it should if the schemes of the non-supersymmetric beta function ansatz and NSVZ beta function coincide, and hence the assumption underlying the matching calculation described here is consistent.

The main conclusion from above is that this third constraint does not provide new information. Rather, any choice for $j$ and $v$ leading to two-loop matching will automatically lead also to the desired matching onto SYM and some uncertainty on the choice of the parameters $j$ and $v$ hence remains. We will consider the following parametrization,
\be
j&=&\frac{34- 10Z\delta}{11-2Z}=\frac{34- 10 Z\delta}{3\beta_0},\nonumber\\ 
v&=&\frac{(1-\delta)10 Z+6C_2(R)Z/N_c}{11-2Z}=\frac{(1-\delta)10 Z+6C_2(R)Z/N_c}{3\beta_0},
\label{parameters}
\ee
so that $\delta=1$ corresponds to the choice introduced in \cite{Ryttov:2007cx}. We will now first discuss how the gravity solution can provide further constraints on these parameters. As we have already stated earlier we do not address the details of adding matter by introducing new dynamical degrees of freedom into the bulk. Rather, we adopt a more exploratory point of view by considering the conjectured beta function as an effective description of the influence of matter on the dilaton field. We have seen that in the supersymmetric case this leads to consistent results.

The conjectured beta-function is expected to describe the gauge theory which for small number of flavors confines, develops a fixed point as the number of flavors is increased and finally looses asymptotic freedom at $Z=5.5$. When approaching this upper boundary of the conformal window from below, $\gamma\ll 1$ and our approximation for taking only the linear term into account should be sufficient. We also know the generic structure of the gravity solution, namely that as we approach the upper boundary of the conformal window from below, there is a fixed point $\lambda^\ast\rightarrow 0$ and the dilaton potential $V(\lambda)$ has a maximum at $\lambda^\ast$. Moreover, there will be a zero of the dilaton potential at $\lambda_0>\lambda^\ast$. As we have seen for the supersymmetric case, this structure should remain throughout the conformal window. Requiring this to be the case also here imposes severe restriction on the parameter $\delta$ in (\ref{parameters}) as we now show.

Recall that for the conjectured beta function
\be
\beta(\lambda)=-2\lambda^2\frac{\beta_0+r\gamma}{1-j\lambda},
\label{beta_again}
\ee
the dilaton potential has zeros at
\be
1-X^2 = 0 = 1- \left(\frac{2\lambda_0(\beta_0+r\gamma)}{3(1-j\lambda_0)}\right)^2,
\label{allorder_scheme}
\ee
which implies that
\be
2\beta_0\lambda_0(\beta_0+r\gamma(\lambda_0))=\pm 3(\beta_0 - \beta_0j\lambda_0).
\ee
Since $r$ is finite as $\beta_0\rightarrow 0$ on the basis of (\ref{parameters}) and (\ref{dict}), the left-hand side of the above equation is zero as $Z\rightarrow 5.5$, i.e. $\beta_0\rightarrow 0$. Therefore, since $\lambda_0\neq 0$, we have $\beta_0 j=0$ which, using (\ref{parameters}) implies that $\delta=34/55$. Hence, the gravity solution we have discussed in this paper seems to fix the parameters $j,v$ uniquely within this framework. Of course the validity of the application of this method to the case at hand can be questioned; nevertheless this result has interesting implications as we shall see.  

We also note the following important property which only holds for $\delta=34/55$. Namely, only for this value the parameter $j=34/11$ independently of $Z$. For any other value of $\delta$ the pole in the beta function (\ref{beta_again}), $1-j\lambda$ will move as more fermion flavors are added to the theory. In particular, as $Z$ is increased the pole will eventually coincide with $\lambda=0$ and spacetime will shrink to a point, which we interpret as unphysical behavior. Note that for SQCD we had $j=2$ independent of the matter fields. 

Let us then turn to the determination of the conformal window.  Let us come back again to Eq. (\ref{beta_again}) and use $\delta=34/55$ to obtain values for parameters $j$ and $v$ from (\ref{parameters}). We may then solve for the lower boundary of the conformal window, $Z^\ast$ by setting the numerator of Eq. (\ref{beta_again}) to zero, $\beta_0(Z^\ast)+r(Z^\ast)\gamma_{\rm{max}}^\ast=0$  with $r$ given by Eq. (\ref{dict}) and $\gamma_{\rm{max}}^\ast$ denoting the critical point value of the anomalous dimension at the lower boundary of the conformal window. The result is
\be
Z^\ast = \frac{11}{2+(1+\frac{7}{11}\frac{N_c}{C_2(R)})\gamma_{\rm{max}}^\ast}.
\label{CW_lowerbound}
\ee

The absence of negative norm states in conformal field theory implies the well-known unitarity bound $\gamma \le 2$ \cite{Mack:1975je,Flato:1983te,Dobrev:1985qv} and this can be used to obtain a lower bound for the conformal window. It turns out that for fixed value of $\gamma^\ast_{\rm{max}}$ the conformal windows are larger when $\delta=34/55$ is used instead the value $\delta=1$ corresponding to the case in \cite{Ryttov:2007cx}. The critical number of flavors obtained using $\delta=34/55$, $\gamma^\ast=0.5$ and the definition of $Z^\ast=2T(R)N_f^\ast/N_c$ are presented in Table \ref{CWtable1} for fundamental, adjoint, and two-index (anti)symmetric fermion representations and different number of colors. 

\begin{table}[htb]
\caption{Lower end of the conformal window assuming $\gamma^\ast=0.5$ for fundamental (F), 2-index (anti)symmetric (2AS) 2S and adjoint (A) representations.}
\label{CWtable1}
\begin{center}
\begin{tabular}{|c|c|c|c|c|c|}
\hline
$N_c$ & $N_{f,{\rm{min}}}$ Fund. & $N_{f,{\rm{min}}}$ 2AS & $N_{f,{\rm{min}}}$ 2S & $N_{f,{\rm{min}}}$ A.\\
\hline
2          & 6.57           & -       &  121/62	     & 121/62 \\	
3          & 10.26         &10.26     &  2.37     & 121/62 \\
4          & 13.84         &7.31      &  2.64   & 121/62 \\
10        & 35.00          &4.81     &  3.28   &  121/62\\
$\to\infty$& $242N_c/69$ &121/31      & 121/31       & 121/62 \\
\hline 
\end{tabular}
\end{center}
\end{table}

\begin{figure}[htb]
\centering
\includegraphics[width=4.0in]{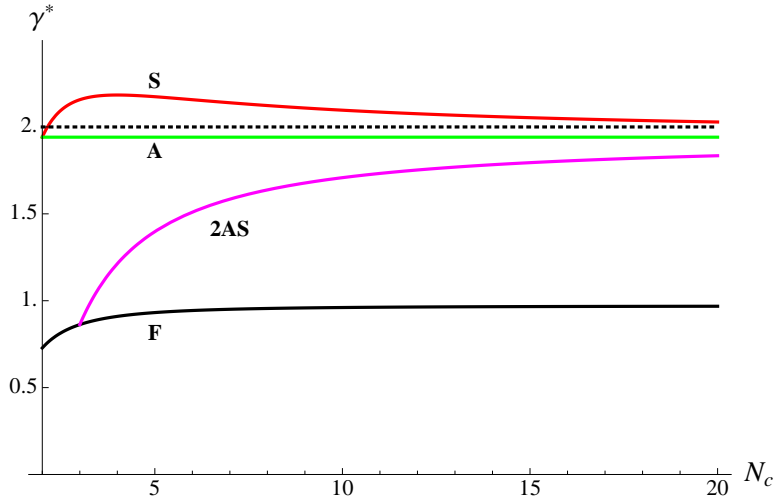}
\caption{Limiting values of $\gamma^\ast$ at the lower boundary of the conformal window if the boundary is required to lie at $\widetilde{Z}^\ast$ determined by $\beta_1(\tilde{Z}^\ast)=0$. The labels F, 2AS, A and S stand for fundamental, two-index antisymmetric, adjoint and two-index symmetric, respectively. The dashed line shows the unitarity limit.}
\label{gammalimit}
\end{figure}

However, we can also proceed using alternative logic. Namely, if we would know the location of the conformal window, i.e. the value $Z^\ast$ for given fermion representation, then we could use (\ref{CW_lowerbound}) to obtain the value of $\gamma_{\rm{max}}^\ast$. As an example, we can consider the lower boundary $\widetilde{Z}^\ast$ determined by $\beta_1(\widetilde{Z}^\ast)=0$ (which we by no means imply to be physical but simply choose as a concrete example). Taking $Z^\ast=\widetilde{Z}^\ast$ in (\ref{CW_lowerbound}) and solving for $\gamma_{\rm{max}}^\ast$ leads to patterns shown in Fig. \ref{gammalimit}. In this special case, as a function of $N_c$, for fundamental representation $\gamma_{\rm{max}}^\ast$ remains always smaller than one ($\gamma^\ast_{\infty}=33/34$), while for adjoint and two index symmetric and antisymmetric representations $\gamma_{\rm{max}}^\ast$ approaches value $\gamma^\ast_{\infty}=33/17$. Interestingly, for two-index symmetric representation $\gamma_{\rm{max}}^\ast$ approaches the same limiting value from above and this is the only representation for which the unitary limit $\gamma\le 2$ provides an useful bound within this example. 

Furthermore, this approach implies two interesting features: first, the value of the anomalous dimension at the lower boundary of the conformal window can be different for different representations, and second, the value for a fixed representation can vary with respect to $N_c$. 

As another example consider the following: if electromagnetic dual theories exist for nonsupersymmetric gauge theories similarly to their supersymmetric counterparts, then the lower bound of the conformal window in electric variables can be determined by determining when asymptotic freedom is lost in terms of magnetic variables. Recently proposals for dual theories for $N_c=3$ nonsupersymmetric gauge theories have appeared both for fundamental \cite{Sannino:2009qc} and higher representations \cite{Sannino:2009me}. For fundamental theories the prediction is that $Z_F^\ast=11/4$ which implies that $\gamma^\ast=88/107\approx 0.82$ at the lower boundary for three colors which was the case for which the dual theory was constructed. Then, for two-index symmetric representation the result obtained from duality is $Z_{2S}^\ast=11/3$ and this leads to $\gamma^\ast=110/173\approx 0.64$. These results again hold for $N_c=3$. It should be noted that in this case the theory with two flavors is not within the conformal window but just below it.

Once the extent of the conformal window has been determined, the beta function ansatz also predicts the values of the anomalous dimension at FP for theories within the conformal window. For example, if the theory with $N_f=2$ flavors in two-index symmetric representation and with $N_c=2,3$ is within the conformal window, then in the case of $\delta=34/55$ we predict that, respectively, $\gamma^\ast=0.46,0.83$. For $N_c=3$ with 9 fundamental flavors we obtain $\gamma^\ast=0.69$ and for 12 fundamental flavors $\gamma^\ast=0.31$.

Although these results still are, more or less, toy examples they provide a concrete setup where the value of the anomalous dimension at the lower boundary of the conformal window can be less than one. This possibility furthermore implies that at fixed points inside the conformal window the value of the anomalous dimension will be even smaller. The lattice simulations are currently starting to investigate these properties in detail. Current results are still inconclusive but future results are certainly expected to shed further light into the dynamics of nonsupersymmetric theories close to or within the conformal window and hence also to the extent to which these analytic beta-function ans\"atze capture the essential features of gauge theories at strong coupling \cite{DeGrand:2009hu,Fodor:2009rb,Shamir:2008pb,Appelquist:2007hu,Hietanen:2009az,Deuzeman:2009mh,Catterall:2007yx,Jin:2009mc,Sinclair:2009ec,DelDebbio:2008zf}.

\section{Conclusions}\label{outlook}

In this paper we analyzed the generic beta function supported by the analytic form of exact NSVZ result in the coupled 5D gravity-dilaton environment. We started from the known results for the beta function in the SYM-SQCD theories and showed that dilaton potential encodes the information about the existence of the poles and possible FPs into its structure. In absence of the FP, we found that the pole in a physical beta function appears as a non-trivial zero in its potential at finite value of the coupling. The presence of the FP in the beta function on the other hand appears as an extremum of the dilaton potential. While this is only a necessary condition, we further confirmed the existence of the FP by studying the flow of the coupling to the constant value at low energies as a function of the domain-wall variable. 

Non-supersymmetric proposals were additionally tested and new insights into the conformal window as well as scheme-independent value of the anomalous dimension at the fixed point were presented. In particular, for the all-orders beta function conjecture of the analytic form first introduced in \cite{Ryttov:2007cx} we determined a parametrization which most closely resembles its supersymmetric exact counterpart; we also presented evidence in favor of these parameter values using holographic gravity duals. It would be extremely interesting to test existing solutions for the dual theories with present method via similar approach as we outlined for the supersymmetric case. 

It would be also interesting to include higher terms in the perturbative expansion of the anomalous dimension in the NSVZ beta function. Including higher powers of $\gamma$ has a special role as it would bring the NSVZ beta function to the form allowing for the FP merger mechanism to occur \cite{Gies:2005as,Kaplan:2009kr}. This also was the core addition to the Ryttov-Sannino beta function proposed in \cite{Antipin:2009wr}. In connection to this construction we can already state some implications that the present work has. The beta function proposed in \cite{Antipin:2009wr} has the form \cite{ATnote}
\be \label{ATbeta}
	\beta (\lambda) &=& - 2\lambda^2
	\frac{\beta_0-r\gamma(\lambda)+s\gamma^2(\lambda)}{1-j\lambda+k\lambda^2}.
\ee
The parameters $r$ and $j$ are fixed on the basis of matching with the two-loop perturbative result as explained in Sec. \ref{RS}, while parameter $k$ is determined by considering the matching onto SYM as explained in \cite{Antipin:2009wr}. For the determination of conformal window the denominator is not needed and in particular the value of $k$ never enters. The calculation carried out in Sec. \ref{RS} to determine the value of the parameter $\delta$ which then fixes $r$ and $j$, can be directly applied also to (\ref{ATbeta}). Hence, also for this ansatz $\delta=34/55$. The essential new ingredient is the use of the holographic relation
\be\label{rooteq}
\gamma_1+\gamma_2=2+\epsilon
\ee
relating the values of the anomalous dimensions at the fixed point obtained by solving for the zeros of the numerator in (\ref{ATbeta}). Here $\epsilon$ represents possible finite $N_c$ corrections; Eq. (\ref{rooteq}) strictly holds only in the large $N_c$ limit. Imposing this constraint on the solutions fixes $s=r/(2+\epsilon)$. Additionally, within this framework, the lower boundary of CW is determined as the point $Z^\ast$ where the two zeros of the numerator in (\ref{ATbeta}) become complex. In particular, at this point $\gamma^\ast=\gamma_1=\gamma_2=1+\epsilon/2$ and in terms of $\epsilon$ the lower boundary of CW is at
\be
Z^\ast = \frac{22}{4+(1+\frac{7}{11}\frac{N_c}{C_2(R)})(1+\frac{\epsilon}{2})}.
\label{CW_lowerboundAT}
\ee
Hence, similar uncertainty as in (\ref{CW_lowerbound}) remains also here: one does not know the precise value of $\gamma^\ast$ at the lower boundary. For positive $\epsilon$ the ansatz (\ref{ATbeta}) leads to larger values of $\gamma^\ast$ than (\ref{beta_again}) for equal values of $Z^\ast$. 

\begin{figure}[htb]
\begin{minipage}{0.5\linewidth}
%\centering
\hspace{-1cm}\includegraphics[width=3.5in]{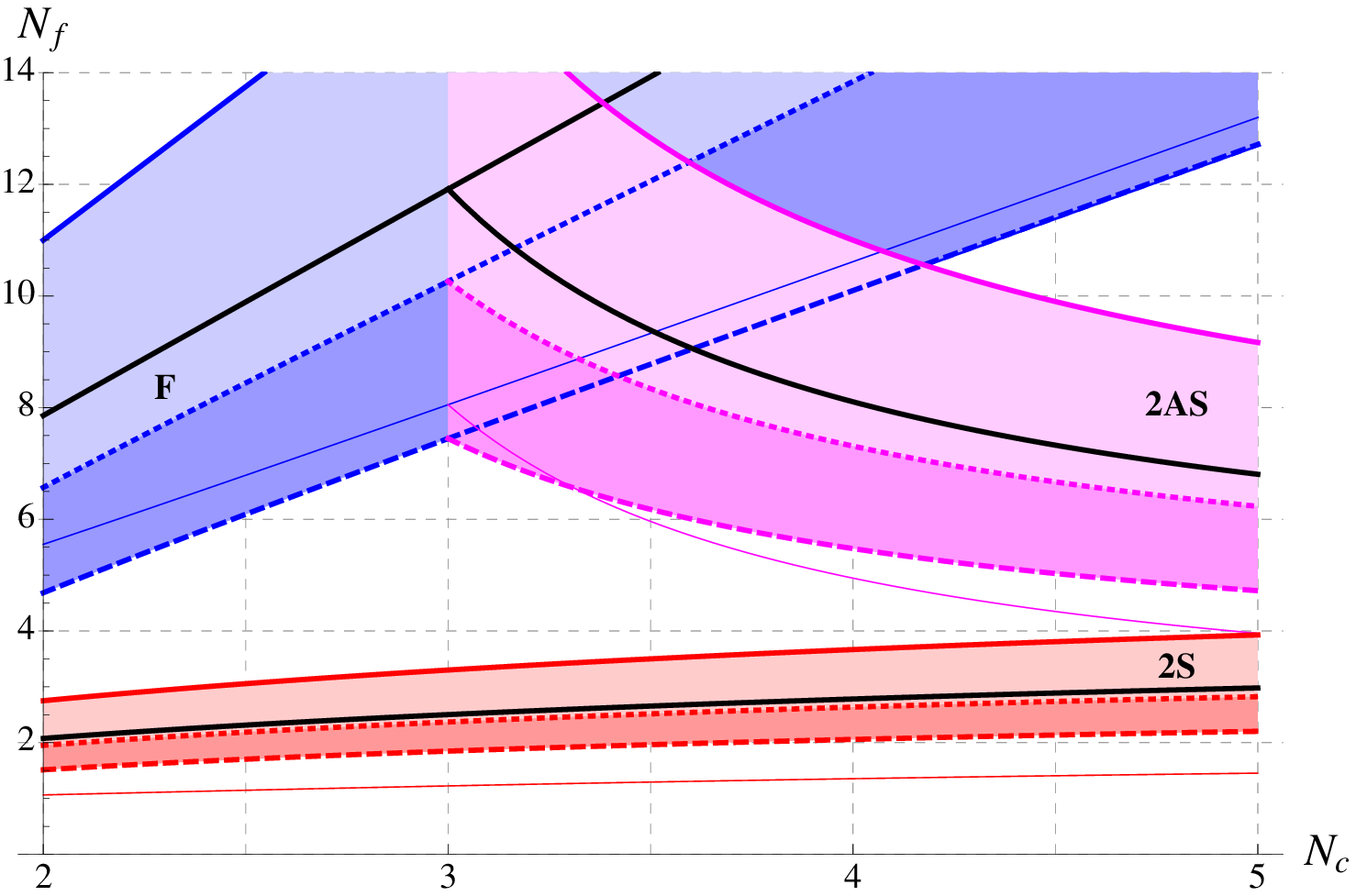} 
\end{minipage}%
\begin{minipage}{0.5\linewidth}
\centering
\includegraphics[width=3.5in]{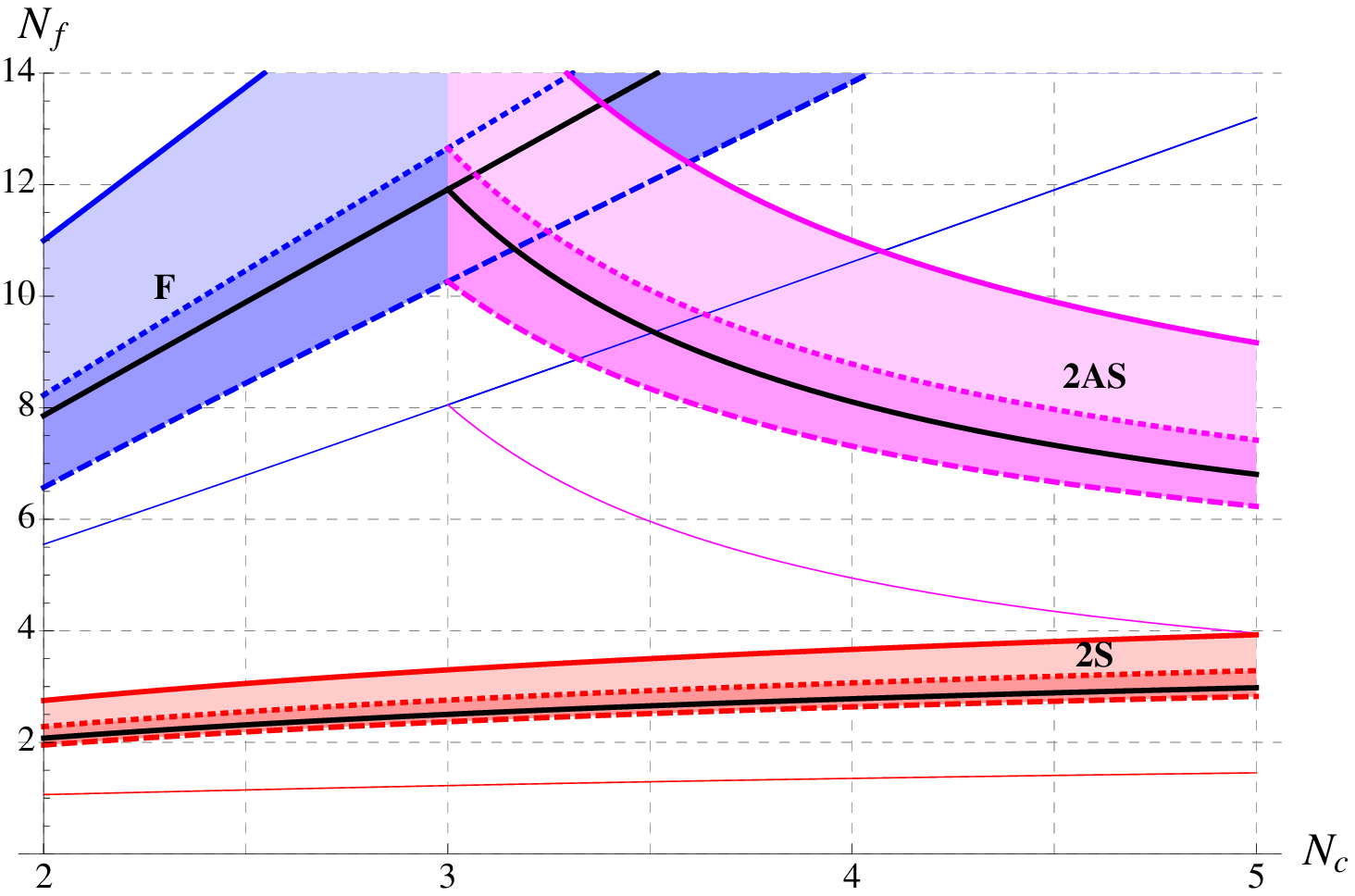} 
\end{minipage}
\caption{Left panel: The phase diagrams obtained using the Ryttov-Sannino like beta function (\ref{beta_again}) with the constraint $\delta=34/55$ on the parameters. Right Panel: The phase diagram for the beta-function (\ref{ATbeta}) with the same constraint. In both figures, for each representation, two different lower boundaries of the conformal window are shown. These correspond to $\gamma^\ast=0.5,1$ (dotted and dashed lines, respectively). Different fermion representations are denoted by F (fundamental), 2S (2AS) (two-index (anti)symmetric). Results for adjoint representation correspond to horizontal lines coinciding with 2S at $N_c=2$. Solid black curves inside the conformal windows in the diagrams correspond to the ladder approximation and the thin colored solid lines correspond to $\beta_1=0$ condition.}
\label{PDs}
\end{figure}

In conclusion, we have presented a holographic analysis which provides further constraints on the beta-function ans\"atze introduced earlier in the literature. However, some uncertainty still remains, related to the value of the anomalous dimension of the quark mass operator at the lower boundary of the conformal window. We summarize the present situation in the two phase diagrams shown in Fig. \ref{PDs}. The left panel shows the conformal windows for various representations obtained using our result $\delta=34/55$ in (\ref{beta_again}) while the right panel shows corresponding results for the beta function ansatz (\ref{ATbeta}). In both figures the dotted and dashed lines denote the lower boundary of CW obtained for $\gamma^\ast=0.5$ and $\gamma^\ast=1$ respectively.

As the figure shows, and as can be directly inferred from formulas (\ref{CW_lowerbound}) and (\ref{CW_lowerboundAT}), the correspondence between the lower limits for CW obtained from these two ans\"atze is that the anomalous dimension at the lower boundary in (\ref{ATbeta}) is twice as large as the one in (\ref{beta_again}). More insight into these beta function ans\"atze is expected through a more complete treatment of fermions in the gravity dual. We hope to return to these questions in a future work.

\acknowledgments
We thank T.\,Alho, K.\,Kajantie and F.\,Sannino for useful and insightful correspondence.


\begin{thebibliography}{199}
\bibitem{jnote}
Note that this is in parallel to pure SYM where $j=\beta_1$/$\beta_0=6/3=2$.
\end{thebibliography}
\end{document}